\PassOptionsToPackage{table,usenames,dvipsnames}{xcolor}
\documentclass[10pt,a4paper,hidelinks]{article}
\newcommand{\mytitle}{Inferring Input Grammars from~Dynamic~Control~Flow}

\usepackage{fancyhdr}
\usepackage{pdfpages}

\fancypagestyle{CISPA}{

\fancyhead[C]{\includegraphics[width=4cm]{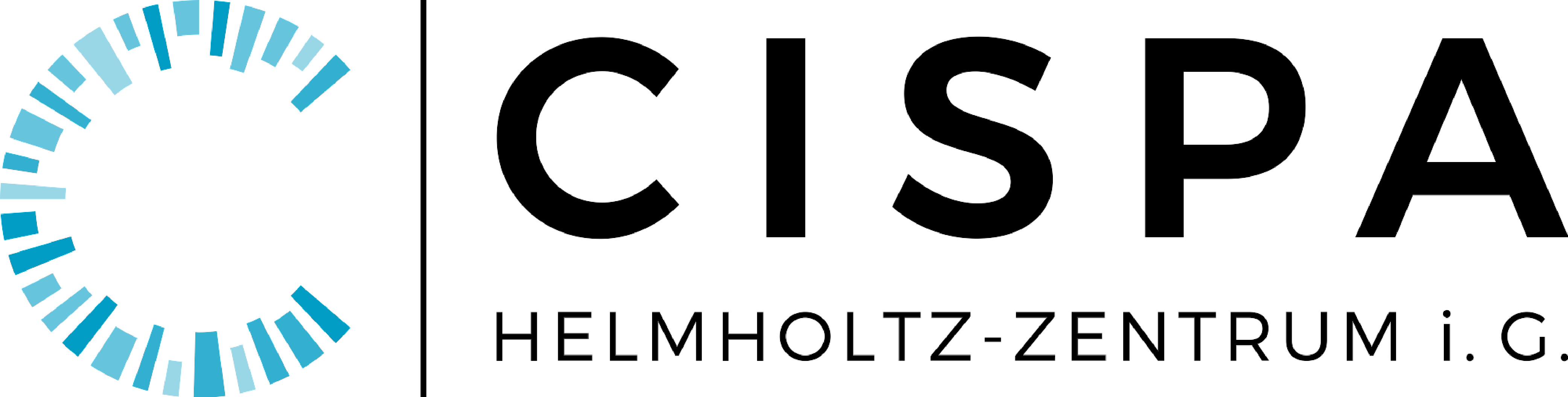}}
}
\usepackage{etoolbox}
\usepackage[backend=bibtex,bibencoding=utf8]{biblatex}
\usepackage{etoolbox}
\newif\ifdraft\draftfalse
\newif\iflong\longtrue

\usepackage[T1]{fontenc}
\usepackage[utf8]{inputenc}
\usepackage{csquotes}
\usepackage[english]{babel}
\usepackage[nohead,includefoot,margin=2cm]{geometry}
\usepackage[compact,raggedright,small]{titlesec}
\usepackage[affil-it]{authblk}
\usepackage[runin]{abstract}
\usepackage{multicol}
\usepackage{titling}
\usepackage{montserrat}
\usepackage{setspace}

\usepackage{multirow}

\usepackage{hyperref}
\hypersetup{colorlinks=true,citecolor=blue,linkcolor=blue,filecolor=blue,urlcolor=blue,pdftitle={\mytitle}}
\usepackage[inline]{enumitem}
\usepackage{stfloats}

\usepackage{color}
\usepackage{relsize} 

\usepackage{tabularx}

\def\BibTeX{{\rm B\kern-.05em{\sc i\kern-.025em b}\kern-.08em
    T\kern-.1667em\lower.7ex\hbox{E}\kern-.125emX}}

\title{Fuzzing with Fast Failure Feedback}

\date{\small (Dated \today)}
\author{Rahul Gopinath}
\author{Bachir Bendrissou}
\author{Bj\"orn Mathis}
\author{Andreas Zeller}
\affil{\{rahul.gopinath, bachir.bendrissou bjoern.mathis, zeller\}@cispa.saarland \\
CISPA - Helmholtz Center for Information Security, Saarbr\"ucken, Germany}

\newcommand\BackgroundPic{
    \put(0,0){
    \parbox[b][\paperheight]{\paperwidth}{%
    \vfill
    \centering
    \includegraphics[width=\paperwidth,height=\paperheight]{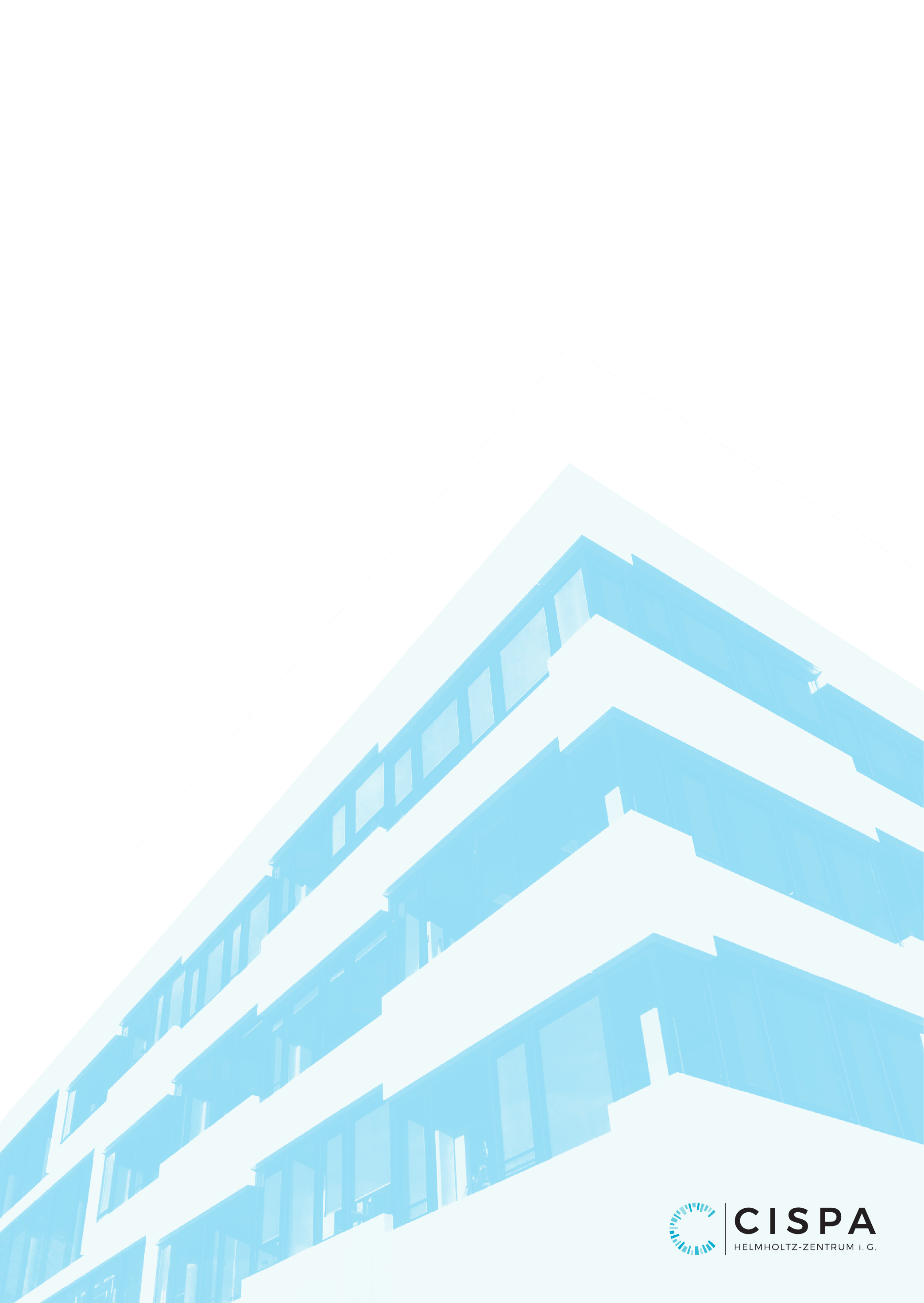}
    \vfill
}}}

\usepackage{utopia}
\usepackage{times}

\usepackage{pgf,tikz}
\usepackage{pifont}
\usepackage{mathtools}
\DeclarePairedDelimiter{\ceil}{\lceil}{\rceil}
\usepackage{doi}
\usepackage{microtype}
\usepackage{dirtree}
\usepackage{pstricks,pst-node,pst-tree}
\usepackage{comment}
\usepackage{amsmath,amssymb,amsfonts}
\usepackage[nameinlink,noabbrev]{cleveref}  

\usepackage{footmisc}
\usepackage{algpseudocode}
\usepackage{algorithm}
\usepackage{algorithmicx}
\usepackage{graphicx}
\usepackage{epstopdf}
\usepackage{textcomp}
\usepackage{xcolor}
\usepackage{xspace}
\usepackage{boxedminipage}
\usepackage{float}
\usepackage{soul}
\usepackage{alltt}

\usepackage{mdwtab}
\usepackage{syntax}
\usepackage[procnames]{listings}
\usepackage{lstautogobble}
\usepackage{qtree}

\definecolor{rltred}{rgb}{0.5,0,0}
\definecolor{rltgreen}{rgb}{0,0.5,0}
\definecolor{rltblue}{rgb}{0,0,0.5}
\hypersetup{
colorlinks=true,
urlcolor=rltblue, 
linkcolor=rltblue,
citecolor=rltblue,
}
\Crefname{figure}{Fig.}{Figs.}
\crefname{section}{Section}{Sections}
\crefname{subsection}{Section}{Sections}
\Crefname{Algorithm}{Alg.}{Algs.}

\usepackage{caption}
\usepackage{subcaption}

\usepackage{balance}
\usetikzlibrary{shapes.misc, positioning}

\usepackage{url}
\usepackage{booktabs}

\usepackage{ntheorem}

\newcommand{\blackbox}{black-box\xspace}
\newcommand{\Blackbox}{Black-box\xspace}
\newcommand{\whitebox}{white-box\xspace}
\newcommand{\Whitebox}{White-box\xspace}

\algblockdefx[switch]{Switch}{EndSwitch}%
[1]{\textbf{switch} #1}%
{}%
\algloopdefx{Case}[1]{\textbf{case} #1:}%
\algdef{SE}[DOWHILE]{Do}{doWhile}{\algorithmicdo}[1]{\algorithmicwhile\ #1}%

\usepackage{contour}
\usepackage[normalem]{ulem}

\contourlength{0.8pt}

\newenvironment{result}%
{\medskip
\noindent
\let\emph=\textbf\begin{center}
\begin{boxedminipage}{\columnwidth}\begin{center}\em}%
{\end{center}\end{boxedminipage}\end{center}%
\medskip
}

\def\term#1{\texttt{'\textbf{#1}'}}
\newcommand{\termC}[2]{\texttt{'\textbf{\color{#2}#1}'}}

\definecolor{nontermcolor}{rgb}{0.4, 0.05, 0.0} 
\definecolor{absnontermcolor}{rgb}{0.4, 0.5, 0.0} 
\definecolor{evokcolor}{rgb}{0.8, 0.05, 0.1}    
\definecolor{incorrectcolor}{rgb}{0.5, 0.0, 0.13}
\definecolor{incompletecolor}{rgb}{0.0, 0.0, 0.55}
\definecolor{validcolor}{rgb}{0.33, 0.42, 0.18} 
\definecolor{retcolor}{rgb}{0.65, 0.16, 0.16}
\definecolor{ipatterncolor}{rgb}{0.0, 0.5, 0.4} 
\definecolor{termcolor}{rgb}{0.0, 0.05, 0.4}    
\definecolor{regexcolor}{rgb}{0.1, 0.3, 0.1}    

\def\term#1{\texttt{\color{termcolor}'\textbf{#1}'}}

\newcommand{\cmark}{\ding{51}}%
\newcommand{\xmark}{\ding{55}}%
\def\Rincomplete{\texttt{\color{incompletecolor}\textbf{\cmark}}\xspace}
\def\Rincorrect{\texttt{\color{incorrectcolor}\textbf{\xmark}}\xspace}
\def\Rvalid{\texttt{\color{validcolor}\textbf{\textcircled{$\checkmark$}}}\xspace}

\renewcommand{\ulitleft}{\normalfont\ttfamily}
\renewcommand{\litleft}{\bgroup\color{termcolor}`\ulitleft}
\renewcommand{\litright}{\ulitright'\egroup}

\renewcommand{\syntleft}{\bgroup\color{nontermcolor}$\langle$\normalfont\itshape}
\renewcommand{\syntright}{$\rangle$\egroup}

\newenvironment{densegrammar}{%
\begin{small}
\begin{grammar}%
}%
{%
\end{grammar}%
\end{small}%
}
{\begin{itemize}\item[]\begin{densegrammar}}%
{\end{densegrammar}\end{itemize}}%

\def\|#1|{\textit{#1}}
\def\<#1>{\texttt{#1}}
\def\[[#1\]]{\texttt{#1}} 

\newcounter{todocounter}
\newcommand{\todo}[1]{\marginpar{$|$}\textcolor{red}{\stepcounter{todocounter}\footnote[\thetodocounter]{\textcolor{red}{\textbf{TODO }}\textit{#1}}}}

\newcommand{\rem}[1]{\textcolor{red}{\textbf{REMOVED }\st{#1}}}

\newcommand{\done}[1]{\textcolor{green}{\stepcounter{todocounter}\endnote[\thetodocounter]{\textbf{DONE }\textit{#1}}}}

\renewcommand{\todo}[1]{}
\renewcommand{\done}[1]{}
\renewcommand{\rem}[1]{}

\definecolor{eclipseBlue}{RGB}{42,0.0,255}
\definecolor{eclipseGreen}{RGB}{63,127,95}
\definecolor{eclipsePurple}{RGB}{127,0,85}

\newcommand{\bFuzzer}{\textsc{bFuzzer}\xspace}
\newcommand{\bFuzzerNS}{\textsc{bFuzzer}\xspace}
\newcommand{\pFuzzer}{pFuzzer\xspace}
\newcommand{\pFuzzerNS}{pFuzzer\xspace}
\newcommand{\lFuzzer}{lFuzzer\xspace}

\definecolor{eclipseBlue}{RGB}{42,0.0,255}
\definecolor{eclipseGreen}{RGB}{63,127,95}
\definecolor{eclipsePurple}{RGB}{127,0,85}

\usepackage[procnames]{listings}
\lstdefinestyle{python}
{
    basicstyle=\footnotesize\ttfamily,
    numberblanklines=false,
    language=python,
    tabsize=2,
    commentstyle=\color{eclipseGreen},
    keywordstyle=\bfseries\color{eclipsePurple},
    stringstyle=\color{eclipseBlue},
    procnamestyle=\bfseries\color{black},
    numbers=left,
    procnamekeys={def},
    columns=flexible,
    xleftmargin=3mm,
    framexleftmargin=1mm,
    identifierstyle=
}

\definecolor{codegreen}{rgb}{0,0.6,0}
\definecolor{codegray}{rgb}{0.5,0.5,0.5}
\definecolor{codepurple}{rgb}{0.58,0,0.82}
\definecolor{backcolour}{rgb}{0.95,0.95,0.92}

\lstdefinestyle{mystyle}{
    basicstyle=\footnotesize\ttfamily,
    commentstyle=\color{codegreen},
    keywordstyle=\color{eclipsePurple},
    escapeinside={(*}{*)},          
    stringstyle=\color{codepurple},
    breakatwhitespace=false,
    breaklines=true,
    captionpos=b,
    keepspaces=true,
    morekeywords={where},            
    showspaces=false,
    showstringspaces=false,
    showtabs=false,
    tabsize=2,
    morestring=[b]' 
}

\algblockdefx[switch]{Switch}{EndSwitch}%
[1]{\textbf{switch} #1}%
{}%
\algloopdefx{Case}[1]{\textbf{case} #1:}%
\algdef{SE}[DOWHILE]{Do}{doWhile}{\algorithmicdo}[1]{\algorithmicwhile\ #1}%

\newcommand{\json}{\textsc{json}\xspace}
\newcommand{\csv}{\textsc{csv}\xspace}
\newcommand{\ini}{\textsc{ini}\xspace}
\newcommand{\tinyc}{\textsc{tinyC}\xspace}
\newcommand{\mjs}{\textsc{mjs}\xspace}

\begin{document}
\AddToShipoutPicture*{\BackgroundPic}

\makeatletter
\renewcommand{\Authfont}{\normalsize\sffamily\bfseries}
\renewcommand{\Affilfont}{\normalsize\sffamily\mdseries}
\begin{titlepage}
\newcommand{\HRule}{\rule{\linewidth}{0.1mm}}
\centering
  \textsc{\LARGE {\fontfamily{Montserrat-TOsF}\selectfont CISPA Helmholtz-Zentrum i.G.}}\\[1.5cm]

  \vspace{2.4 cm}
  \HRule \\[0.2cm]
  {\huge\sffamily\bfseries \@title\par}
  \vspace{0.2cm}
  \HRule \\[1.5cm]

  {\sffamily \@author\par}
\vfill

\end{titlepage}

\makeatother
\setlength{\affilsep}{0.1em}
\addto{\Affilfont}{\small}
\renewcommand{\Authfont}{\normalsize}
\renewcommand{\Affilfont}{\normalsize}

\pretitle{\begin{center}\large\bfseries}
\posttitle{\end{center}}
\maketitle
\thispagestyle{CISPA}

\begin{abstract}
Fuzzing---testing programs with random inputs---has become the prime technique to detect bugs and vulnerabilities in programs.
To generate inputs that cover new functionality,
fuzzers require \emph{execution feedback} from the program---for instance, the coverage obtained by previous inputs, or the conditions that need to be resolved to cover new branches.
If such execution feedback is not available, though, fuzzing can only rely on chance, which is ineffective.
In this paper, we introduce a novel fuzzing technique that relies on \emph{failure feedback} only---that is, information on whether an input is valid or not, and if not, where the error occurred.
Our \emph{\bFuzzer} tool enumerates byte after byte of the input space and tests the program until it finds valid prefixes, and continues exploration from these prefixes.
Since no instrumentation or execution feedback is required, \bFuzzer is language agnostic and the required tests execute very quickly.
We evaluate our technique on five subjects, and show that \bFuzzer is effective and efficient even in comparison to its \whitebox counterpart.

\end{abstract}
\begin{multicols}{2}

\section{Introduction}
A fuzzer quickly generates artificial inputs for a program under test. 
A fuzzer is \emph{effective} if it can explore sufficiently many regions of the input space of
the program~\cite{boehme2020fuzzing} and implementation~\cite{havrikov2019systematically}
and is \emph{efficient} if it can generate and evaluate
inputs quickly~\cite{boehme2020fuzzing}.
Fuzzing a system requires relatively little
knowledge about the program or domain, 
and has become one of the most popular software bug finding techniques. 

The classical \blackbox fuzzing strategy is to generate lots of
random inputs quickly, and try them one at a time against the program. While
effective in quickly covering shallow code paths and identifying problems in
parsing, this strategy fails when it comes to programs that require complex
structured inputs.
The problem is that for such programs, one needs to get past the parser
by producing \emph{syntactically valid} inputs
to reach the semantics of the program.
This is especially hard to do using a \blackbox approach.
For example, the chance of producing a simple JSON fragment such as \term{[\{\}]} completely
randomly is $1 : 256^4$ or \emph{one in four billion} even when the
size is restricted to just four bytes.
Hence, deeper code paths and program internals of programs with complex input specifications
are out of reach for traditional \blackbox fuzzing.

One can generate valid inputs for a program if one starts with the input grammar
specification of the program, and produce syntactically conforming inputs.
However, input specifications are hard to come by,
and even when available,
may not describe the implementation correctly --- which is precisely where bugs hide.
For example, even for simple formats such as JSON, there are few parsers that
implement it correctly~\cite{seriot2016parsing}.

An alternative is to start with a sample set of \emph{valid inputs,} and rely on
\emph{mutations} of these inputs for generating more inputs.  The problem with this
approach is that it biases the fuzzer towards inputs that are present in the
neighborhood of the sample set. This reduces the effectiveness of the
fuzzer. Further, generating valid inputs even with mutation of valid samples
can be hard. Hence, these fuzzers commonly rely on \emph{coverage feedback} from the
program, with the program running under instrumentation (i.e a
\emph{grey box} technique).
However, 
there can be situations in which 
(1) the system \emph{cannot be instrumented}---for instance, because the system is written in a different programming language
than what the instrumentation expects,
or (2) the system code comes in \emph{read-only memory} (or otherwise cannot be changed), making instrumentation impossible, or
(3) the system is \emph{remote,} and can only be accessed through its interface.
In all these situations, a \blackbox approach is needed.

In this paper, we propose a novel strategy for \emph{quickly generating valid
inputs for \blackbox programs with structured inputs.} Our strategy is based on two
key observations. (1)~Programs with complex input specifications often report
failure as soon as an unexpected input symbol (a byte or a token) is
present in the input.\footnote{This is the same observation made in \pFuzzer from
Mathis et al.~\cite{mathis2019parser},
for \emph{\whitebox} fuzzing, and we will be using \pFuzzer as the baseline for
our evaluations.}
Even if the program is unable to immediately signal failure to parse or
validate, they can often precisely indicate the
maximal \emph{valid prefix} of the input that if combined with some valid
suffix will be accepted by the program. (2)~Once the program can signal
where a failure occurred, there are only a limited number of alternatives
that can be used at that point, and \emph{it is feasible to check them all}.

We use this \emph{failure feedback} to generate continuations that will
ultimately result in valid inputs. Our approach works on \emph{all programs that
satisfy the following conditions}:
(1)~The program should accurately identify and accept valid inputs and reject invalid inputs.
(2)~The program should distinguish between inputs that are merely \emph{incomplete}
(that is, there exists some suffix that will make this input valid)
and those that are \emph{incorrect}
(such a suffix does not exist).
In some cases, such as reading in a chunk of data before validating,
a program may not be able to immediately identify whether an
input is incomplete or incorrect. In such cases, we can also relax the second
constraint to: (2) The program, when it identifies an \emph{incorrect} input,
should report the \emph{maximal valid prefix} that was parsed by the program.
That is, some actually \emph{incorrect} inputs may be identified as
\emph{incomplete} so long as when the program, when it finally realizes that the
input prefix is \emph{incorrect}, precisely identifies where the failure
occurred.

Neither of these conditions are onerous for parsers.
Most parsers already incorporate\footnote{This feature is taken advantage of
by editors such as \emph{vim} and \emph{emacs} to precisely jump to the location
of syntax error.} these conditions as a prerequisite to
help the users in fixing failures.

Given a program that can distinguish between \emph{valid}, \emph{incomplete},
and \emph{incorrect} inputs, our strategy first produces the smallest unit input
symbol accepted by the program (bits, bytes, words, or even tokens if we have
further information on what is expected). This input
will be typically rejected by the program as either \emph{incorrect} or
\emph{incomplete}. If it is rejected as \emph{incorrect}, the fuzzer will
enumerate all possible unit symbols, and try each as a fix, one of which
will be rejected by the program as merely \emph{incomplete}. The fuzzer will
again enumerate all possible unit symbols as a possible extension to the
incomplete input. This process is repeated until either the program accepts
the input or we run out of testing budget for a single item.

If on the other hand, the program is unable to distinguish \emph{incomplete}
and \emph{incorrect} inputs immediately (without reading in further data) but
provides precise feedback on where the failure occurred, we trim the input to
that point, and continue generating possible continuations from that point.

In either case, we only have a fixed number of symbols to verify at each point.
That is, in the first case, if we are using bytes as the alphabet, one only
needs to try $|\alpha| = 256$ values at each point, requiring
only $|\alpha|\times L$ executions to generate an input that is $L$ bytes
long. In the second case, we only have to try a maximum of $|\alpha| \times C$ inputs
at each point
where $C$ is the chunk size that the program needs to read in, resulting in
$|\alpha| \times\frac{L}{C} \times \sum{C}$
executions of the program for generating an input
$L$ long when compared to $|\alpha|^L$ executions if we were reliant on traditional
\blackbox fuzzing.

Since our approach is a \blackbox variation of the \emph{parser directed
fuzzing} used by \pFuzzer, we call our prototype \bFuzzer.
Our approach is similar to, and was inspired by
\pFuzzer~\cite{mathis2019parser}. \pFuzzer is a \whitebox technique, and
uses dynamically tracked byte comparisons within the instrumented program to
decide which next byte to append to the input. \bFuzzer, instead, lets
the program run uninstrumented,
and simply tries all $2^8$ bytes and use any byte that lets it proceed.
Why not simply use a \whitebox approach like \pFuzzer does? Indeed, \whitebox
approaches, where applicable, can be more effective than any \blackbox
approach. However, the realities of software development in the real world
is such that \whitebox techniques cannot always be applied.
\done{You are repeating arguments already brought forward in the same section. Maybe move them to the next section? -- AZ; Moving these to the end -- RG}
\done{Too many arguments at this point. Again, move them to the next section, maybe illustrated by the example? -- AZ; Moved it to the end -- RG}
For example, consider \emph{cyber-physical} systems such as those used by
the automotive industry that are composed of multiple components. If one wants
to instrument such systems, it may be necessary to instrument several
components many of which may be third party with different operating systems
(some of which may not even have operating systems),
with severe constraints on the resources the component is allowed to consume.
For example, a constraint
on the memory allowed may restrict low level instrumentation. Similarly,
a constraint on real-time behavior may restrict the time available for
instrumentation.

While it may be possible to build specialized hardware that can allow
instrumentation with limited impact, it is far easier to adapt return
values to add more information about why an input is invalid (if 
such information is not already returned).



Interestingly, our \bFuzzer approach is competitive with \pFuzzer on
the same subject programs from the \pFuzzer paper~\cite{mathis2019parser}.
\bFuzzer obtains better coverage than \pFuzzer on four
out of five programs tested (\ini, \json, \tinyc and \mjs) while
generating a \emph{much larger number} of valid inputs. These inputs are also
many times larger than the inputs produced by \pFuzzer, which translates to
larger input variety.

Our research has three key contributions:
\begin{itemize}
  \item Our approach is the first \emph{feasible} \blackbox approach targeting
    programs with complex input specifications.
  \item We pioneer using the program response in the form of precise information
    as to where the failure occurred for \blackbox fuzzing.
  \item We identify a robust method of dealing with imprecision in failure
    report: \emph{backtracking} strategies depending on the
    kind of information presented, and the pattern expected.
\end{itemize}

We illustrate our approach with a detailed example in \Cref{sec:example},
and the algorithms involved are detailed in \Cref{sec:approach}.
We conduct an evaluation of our approach in \Cref{sec:evaluation}, explain the
threats to the validity of our experiments in \Cref{sec:threats}, explore the
background and competing research in \Cref{sec:related}, and the limitations in
\Cref{sec:limitations}. \Cref{sec:conclusion} concludes.

\section{Fast Failure Feedback Fuzzing in Action}
\label{sec:example}

\done{Do not name this nutshell if this is the one place in which the approach is defined. Nutshell is for short summaries (to be detailed later) -- AZ}

Consider the following scenario: You are given a program~$P$ whose behavior you want to explore.
For that, you need to generate plausible inputs that quickly cover its input
space, exercising all features of the input language.
You are assured that it is a well behaved program with the following
characteristics.

\begin{figure}[H]
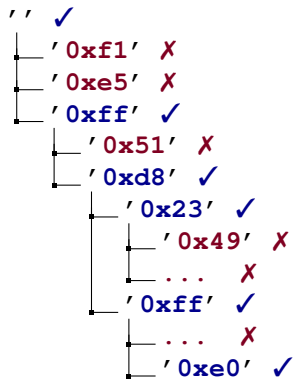

\dirtree{%
.1 \termC{}{incompletecolor} \Rincomplete.
.2 \termC{0xf1}{incorrectcolor} \Rincorrect.
.2 \termC{0xe5}{incorrectcolor} \Rincorrect.
.2 \termC{0xff}{incompletecolor} \Rincomplete.
.3 \termC{0x51}{incorrectcolor} \Rincorrect.
.3 \termC{0xd8}{incompletecolor} \Rincomplete.
.4 \termC{0x23}{incompletecolor} \Rincomplete.
.5 \termC{0x49}{incorrectcolor} \Rincorrect.
.5 \textcolor{incorrectcolor}{\texttt{...}} \Rincorrect.
.4 \termC{0xff}{incompletecolor} \Rincomplete.
.5 \textcolor{incorrectcolor}{\texttt{...}} \Rincorrect.
.5 \termC{0xe0}{incompletecolor} \Rincomplete.
}
  \caption{The process of fast failure exploration resulting in \term{0xff 0xd8 0xff 0xe0}.}
\label{fig:fff}
\end{figure}

\begin{enumerate}
  \item [1.] $P$ processes its input bytes sequentially, and as soon as the
    input is determined to be \emph{invalid} or \emph{valid}, this information
    is conveyed to the user\footnote{In
    POSIX systems, this is easily achieved through the \emph{exit-code} ---
    an 8~bit integer, and equivalents exist in other systems.}.
    The information returned can be one of the following:
  \done{What is a code? State that it is one of the following. -- AZ}
  \item [2.] $P$ returns \Rincorrect if an unexpected byte was found in the input.
  \item [3.] $P$ returns \Rincomplete if the input was a valid prefix, but incomplete.
  \item [4.] $P$ returns \Rvalid if the input was valid (valid prefix and complete).
\end{enumerate}
Given no further information, how do we elicit valid inputs from
such a program?

We start with an empty string --- \term{} as input. This is immediately
rejected by the program as invalid. The program, however, returned \Rincomplete.

Since a return value \Rincomplete indicates that more characters are needed,
we add a byte to the empty string --- say \term{0xf1}.
This input is again rejected by the program. However, this time, it returned
a value \Rincorrect.

The return value \Rincorrect tells us that the program parser encountered an
unexpected input.
Since we have only a single byte in the input string, we replace
that byte with another, returning \Rincorrect.
We repeat the process with other bytes until we try \term{0xff}, at which point
the program responds with \Rincomplete.

As this means that our starting input \term{0xff} was a valid prefix, we
continue by appending one more byte,
generating \term{0xff 0x51}. This again is rejected
with \Rincorrect.

At this point, we know that the initial character is a valid prefix.
That is, the only plausible cause for failure is the newly added character.
Hence, we replace the newly added \term{0x51} with another random byte resulting
in \term{0xff 0xd8} which results in a return value \Rincomplete.

We now have two bytes that are accepted as a valid prefix by the program. Assuming the program
accepts the contents of a file as an input, this is actually sufficient for
guessing the file type. The mystery program accepts files of type
JPEG~\cite{wikipedia2020list}.

For the sake of exposition, let us relax our constraints on the program for a
bit. Rather than assuming immediate feedback, what happens if we get the failure
feedback a little later?

Continuing from before, as the program suggests, we add another random
byte~\term{0x23}, and program responds with \Rincomplete, we add another
byte~\term{0x49}, resulting in the string \term{0xff 0xd8 0x23 0x49}.
At this point, the program returns \Rincorrect.
As before, we start replacing the last added byte with another.
However, this time, none of the possible bytes could elicit any return value
other than \Rincorrect.

Here, we do something different. We know that none of the bytes could continue
the string \term{0xff 0xd8 0x23}. Hence, we go back one step, and replace
\term{0x23} with another byte, say \term{0xff} resulting in \Rincomplete,
continuing again with random bytes, we now find that \term{0xff 0xd8 0xff}
can be continued with \term{0xe0} yielding \Rincomplete allowing us to continue
further.
Note that at this point, we have already generated a valid prefix with four
bytes \term{0xff 0xd8 0xff 0xe0}, confirming our guess that
the mystery program is a JPEG processor.
A graphical representation of the exploration of JPEG parser is given in \Cref{fig:fff}.

Why backtrack? Consider this fragment below:
\begin{lstlisting}[style=python]
data = file.read(2)
if data == [0xff, 0xd8]:   return SOI
elif data == [0xff, 0xc0]: return SOF0
elif data == [0xff, 0xc2]: return SOF2
elif data == [0xff, 0xc4]: return DHT
elif data == [0xff, 0xdB]: return DQT
elif data == [0xff, 0xdd]: return DRI
elif data == [0xff, 0xda]: return SOS
elif data == [0xff, 0xd0]: return RST0
elif data == [0xff, 0xe0]: return APP0
elif data == [0xff, 0xd9]: return EOI
else: raise Fail()
\end{lstlisting}
Here, the failure feedback is deferred. That is,
we will keep receiving \Rincomplete until both bytes have been read, at which
point, the program can return \Rincorrect if the bytes read were not in the
expected set.
Hence, we need to backtrack
the previous bytes where we got \Rincomplete. Since the fuzzer does not know how
many bytes were read in a chunk, we start by backtracking one byte, then two
bytes etc.

There is another possibility. Consider the fragment below:
\begin{lstlisting}[style=python]
idx = file.tell()
data = file.read(2)
if data == [0xff, 0xd8]:   return SOI
...
elif data == [0xff, 0xd9]: return EOI
else:
    raise Fail(till=idx)
\end{lstlisting}
It raises the exception \<Fail(0)> when given an input \term{0xff~0xe1}.
In this case,
we get an \Rincorrect, but with additional information --- the failure occurred
at the token starting at \<0>.
However, the additional information is likely an overapproximation. In this
case, the actual maximal valid prefix is \term{0xff} even though the failure
was signaled at index \<0> (instead of \<1>). If we could trust the additional
information, we would have marked \term{0xff} at index \<0> as having resulted in
failure, and hence, no further data will be generated starting with \term{0xff}.
Instead, when we have additional information from the program during rejection
of incorrect inputs, we may assume that it is an overapproximation within a
certain bound (configurable, but set to a single byte by default,
which assumes no overapproximation).
We then behave as if the index at
which the failure was reported can have two bytes ($256^2$) before we exhaust
it.

A common case is when the format is length-delimited (as many fields in
JPEG are). For example, see the below fragment from \<nanojpeg.py>.
\begin{lstlisting}[style=python]
def njDecode16(pos):
    return (nj.spos[pos]<<8) | nj.spos[pos+1]

def njDecodeLength():
    if (nj.size < 2): raise Incomplete()
    nj.length = njDecode16(nj.pos)
    if (nj.length>nj.size): raise Incomplete()
\end{lstlisting}
The check at \<L5> makes the program return \Rincomplete until the first
two bytes are provided as input. These bytes are interpreted as the length of
the next field (\<L6>). Next, the program returns \Rincomplete until
there is enough data to fill the field.
Here, the length field is fixed to two bytes, and any two random bytes
are sufficient to populate it. Once the field is populated, the bytes in 
that field is interpreted as the length of the next field, and the program
returns \Rincomplete for as many bytes as specified.
This works
fairly well so long as there are few such fields (filling such a field
can take $2^{16}$ executions at worst if two bytes are used to specify the
length), and sufficient memory is available.

A final case is when there is a lexer in front of the parser. Consider the
fragment from \tinyc below.
\begin{lstlisting}[style=python]
pos = lexer.pos()
token = lexer.next()
if token == DO:
  statement()
  token = lexer.next()
  assert token == WHILE
  paren_expr()
\end{lstlisting}
There are two problems here. The first is that, without additional steps,
the fuzzer will only get the failure feedback at the end of completing a
token, at which point, if incorrect, it may be asked to restart. When the token length
is more than two bytes, it becomes expensive to generate tokens this way.
One way to fix this problem is to
incorporate failure feedback into the lexer. We recommend using a
\emph{trie}~\cite{brass2008advanced} that is primed with the tokens
accepted by the lexer. Using a \emph{trie}, we can find the precise location
where token match failed, which is then returned to the user. This allows
one to construct $L$ length tokens using at most $|\alpha| \times L$ executions
rather than $|\alpha|^{L}$ executions in the worst case where $|\alpha|$
is the number of alphabets in the input language --- 256 for a byte. 
This leaves us with the second problem, which is that
the \emph{lexer} does not have the information as to what is the legal
token at each point.
That is, after processing the \<statement()>, the lexer will still
accept \<DO> or \<IF> or other tokens. Here, we use backtracking. However,
rather than trying $|\alpha|^L$ executions, we only have to try at most
$|\alpha|\times L \times |T|$ executions where $|T|$ is the number of tokens
that the lexer knows about, and $L$ is the length of the longest token.


There are a few more heuristics that we have not mentioned so far. For example, if we
backtrack two bytes, and find that we still have a failure, rather than
continuing the same procedure, it may pay better to shift to a dictionary based
approach. We can also enhance the fuzzer with any further information given
by the program, for example, the expected token at any point.
We note that even when backtracking, 
we are always at least as good as simple random fuzzing.

\section{Approach}
\label{sec:approach}
\done{This is not the implementation, but the approach :-) -- AZ; I think it's both -- BM; Renamed it -- RG}

\begin{figure}
\begin{lstlisting}[style=python]
def generate(validate, prefix=''):
    alphabet = [i for i in range(CMAX)]
    seen_at = list()
    seen = set()
    while True:
        choices = [i for i in alphabet
                     if i not in seen]
        if not choices:
            seen, prefix, choices, seen_at =\
              backtrack(prefix, alphabet, seen_at)
        byte = choose_symbol(choices)
        n_prefix = prefix + byte
        rv, n = validate(n_prefix)
        if rv == Complete:
            return n_prefix
        elif rv == Incomplete:
            seen.add(byte)
            prefix = n_prefix
            seen_at.append(seen)
            seen = set()
            alphabet = [i for i in range(CMAX)]
        elif rv == Incorrect:
            if n is None:
                seen.add(byte)
            else:
                if n < len(seen_at):
                    seen = seen_at[n]
                    seen_at = seen_at[:n]
                seen.add(byte)
                rs = len(n_prefix) - n
                alphabet = symbols(range(CMAX),
                               min(rs, OAPPROX))
                prefix = prefix[:n]
\end{lstlisting}
  \caption{The \<generate> function}
\label{fig:gen}
\end{figure}
The core of the algorithm is the \<generate>
algorithm that accepts two parameters: The \<validate> parameter which is the
external validator for the input, and \<prefix> which is an optional prefix
for the input to start from.

We begin by initializing the alphabet (\<L2>). These are the only options we
have for continuation when given a prefix.
We have used \<CMAX> set to 256 to represent all bytes as the basic alphabet.
However, this can be modified by the user to either choose a restricted subset,
or even completely different set of alphabets such as tokens.

Next, we do the following repeatedly until we produce a valid input:
We first produce a set of \<choices> for adding to the current prefix (\<L5>).
These are the \emph{remaining} symbols from \<alphabet> after discounting
any symbols already checked. We choose one symbol, and produce
a new prefix \<n_prefix>, which is used by the \<validate> function. The
return value from validation contains
information about the parse status of the newly created prefix.
Based on the return value, we proceed as follows:
\begin{description}
  \item[Complete:] If the return value is \<Complete>, then exploration ends (\<L15>).
  \item[Incomplete:] If the return is \<Incomplete>, we are ready to add more bytes.
    We mark the current symbol as having been seen in this position by adding
    it to \<seen>, and update the prefix to the value of new prefix (\<L18>),
    update the seen values in the global list of seen values at each index for the
    current exploration sequence (\<L19>), and reinitialize the \<alphabet>. The
    reinitialization of \<alphabet> is necessary in case we find a numeric
    failure index as we will see later.
  \item[Incorrect:]
    With \<Incorrect>, there are two possibilities. (1) When
    there is no number indicating the failure index, it simply means that the
    last symbol added was incorrect. In this case, we simply note this symbol
    as having been seen at this index (\<L24>) and continue. The next iteration
    will replace this symbol, and will never use this symbol with this prefix
    again. (2) When there is a failure index, it means that the failure was
    signaled starting \emph{at least} at that index (the index may be larger).
    In this case, we first check, and get the \<seen> value from that index.
    Next, there is a chance that the failure was signaled at a few symbols
    earlier than the actual failure. This is controlled by the value
    \<OAPPROX>. We recommend a maximum \emph{overapproximation} of 2
    symbols to avoid unbounded exponential growth\done{"to avoid exponential growth" or something similar to shortly explain the recommendation -- BM}
    (and should be set to 1 ideally). That is, if \<OAPPROX> is set
    to \<2>, we will check all one and two symbol sequences \emph{at this
    position} for possible fixes (\<L32>), which gets added to \<seen>.
    The implementation for generating all such symbols is given in \Cref{fig:nsymbol}.
    This is the reason for resetting the symbol set in \<L21>.
\end{description}
\begin{figure}[H]
\begin{lstlisting}[style=python]
def symbols(base_sym, rs):
    alphabets = []
    for r in range(1, rs+1):
        v = [bytes(i) for i in
           product(range(base_sym), repeat=r)]
        alphabets.extend(v)
    return alphabets
\end{lstlisting}
  \caption{Generating symbol sequences}
  \label{fig:nsymbol}
\end{figure}
When multiple iterations happen, it may come to a point that no symbol in the
alphabet is left to be seen at a particular index. When that happens, it is an
indication that we need to backtrack as we mentioned previously. The
backtracking implementation is given in \Cref{fig:backtrack}.
\begin{figure}[H]
\begin{lstlisting}[style=python]
def backtrack(prefix, alphabet, seen_at):
    if not prefix: raise LastIndexException()
    seen = seen_at[len(prefix)-1]
    seen_at = seen_at[:-1]
    prefix = prefix[:-1]
    choices = [i for i in alphabet
                 if i not in seen]
    if not choices:
        return backtrack(prefix, alphabet)
    return seen, prefix, choices, seen_at
\end{lstlisting}
  \caption{The backtracking function}
  \label{fig:backtrack}
\end{figure}
The final piece of the puzzle is how the program should work. \Cref{fig:hello}
is a simple program \<ishello> that checks whether the given string is \<hello>.
It demonstrates how \Rincomplete is returned whenever
the length of \<input> (the \emph{valid prefix})
is less than five bytes (\<L12> \& \<L13>), Similarly \Rincorrect is returned whenever the byte
at a given index is incorrect (\<L3> - \<L8>). It also demonstrates string comparison
necessitating backtracking (\<L9> \& \<L10>). Finally \Rvalid is returned when the complete input matches
\<HELLO>.
\begin{figure}[H]
\begin{lstlisting}[style=python]
def ishello(input):
    try:
        if input[0] != b'H':
            return Incorrect
        if input[1] != b'E':
            return Incorrect
        if input[2] != b'L':
            return Incorrect
        if input[3:5] != b'LO':
            return Incorrect
        return Complete
    except IndexError:
        return Incomplete
\end{lstlisting}
  \caption{An example program for checking if a string is \<hello> that can be fuzzed by
  \bFuzzer.}
  \label{fig:hello}
\end{figure}
\section{Evaluation}
\label{sec:evaluation}

Our technique uses just a bit more information than random fuzzing. Hence,
comparisons should ideally be with respect to pure random fuzzing. However,
we do not have to actually run the fuzzers to determine how they
will compare. Pure random fuzzing has only one in $|\alpha|^L$ chance of
producing a string that is $L$ bytes long where $\alpha$ represents the
alphabet~\cite{godefroid2007random}. Indeed, this is true for any
fuzzer that does not or cannot rely on feedback from program, as they have
to fallback on random chance to generate such strings. Since there is
no other fuzzer that exploits fast failure feedback, \bFuzzer is
superior to other \blackbox fuzzers by construction.
\done{State that this also applies to fuzzers that do not get feedback from the program, and that there is no other fuzzer that exploits fast failure feedback.  (In this way, \bFuzzer is superior by construction already.)~-- AZ}

Hence, for a more relevant and useful comparison, we use
\pFuzzer~\cite{mathis2019parser} as a baseline. Note that while \pFuzzer is
quite similar to \bFuzzer, \pFuzzer is a \emph{\whitebox} technique while \bFuzzer is a
\blackbox technique. That is, \pFuzzer operates with quite a lot more
information than \bFuzzerNS. Further, \pFuzzer was shown to be better than the
state of the art~\cite{mathis2019parser} such as AFL and KLEE~\cite{cadar2008klee} in generating
valid inputs for programs quickly. Is this advantage preserved when we eschew instrumentation?

We use five different programs with complex input language. These are the same
programs that were used by \pFuzzer for its evaluation~\cite{mathis2019parser}, shown in 
\Cref{tbl:subjects}. We modified each slightly so that they would provide
fast failure feedback as required by \bFuzzer.\footnote{Note that our claim is only
applicable to programs that fulfil our criteria---availability of precise
failure reports that distinguish between \Rincomplete and \Rincorrect. We had to
modify these programs so that they fulfil the expected criteria. If the program
satisfies our criteria, we do not require the source code.} The number of lines
added is listed in \Cref{tbl:subjects}. Note that the
modifications of \tinyc and \json include lines for checking whether the
\emph{trie} contains a given token. The implementation of \emph{trie} is in a
shared file (53 lines) and is not included in this count.

\todo{VSpace violation caused by images. -- BM}
\begin{table}[H]
\caption{Subject Programs}
\centering
\begin{tabular}{|l|c|r|c|r|}
  \hline
Name & Accessed & Lines of Code & Kind & +Lines \\
\hline
  \ini   & 2018-10-25 & 293    & RG & 5 \\
  \csv   & 2018-10-25 & 297    & RG & 1\\
  \json  & 2018-10-25 & 2,483  & CFG & 61 \\
  \tinyc & 2018-10-25 & 191    & CFG & 60 \\
  \mjs   & 2018-10-25 & 10,920 & CFG & 6 \\
\hline
\end{tabular}
\label{tbl:subjects}
\end{table}
\begin{figure}[H]
  \centering
  \begin{subfigure}[b]{0.35\textwidth}
  \includegraphics[width=\textwidth]{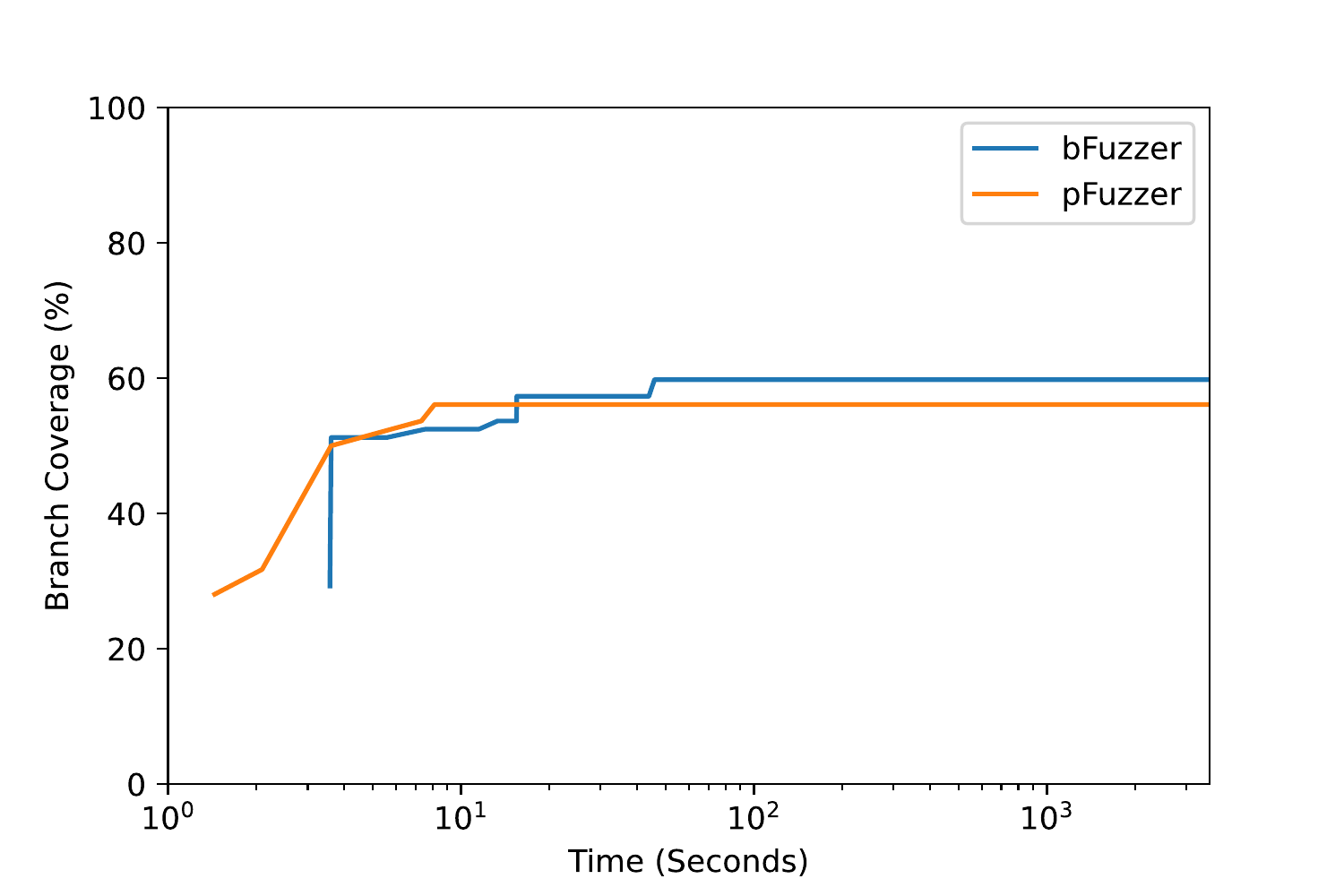}
  \caption{Branch coverage comparison for INI}
  \label{fig:inib}
  \end{subfigure}

  \begin{subfigure}[b]{0.35\textwidth}
  \includegraphics[width=\textwidth]{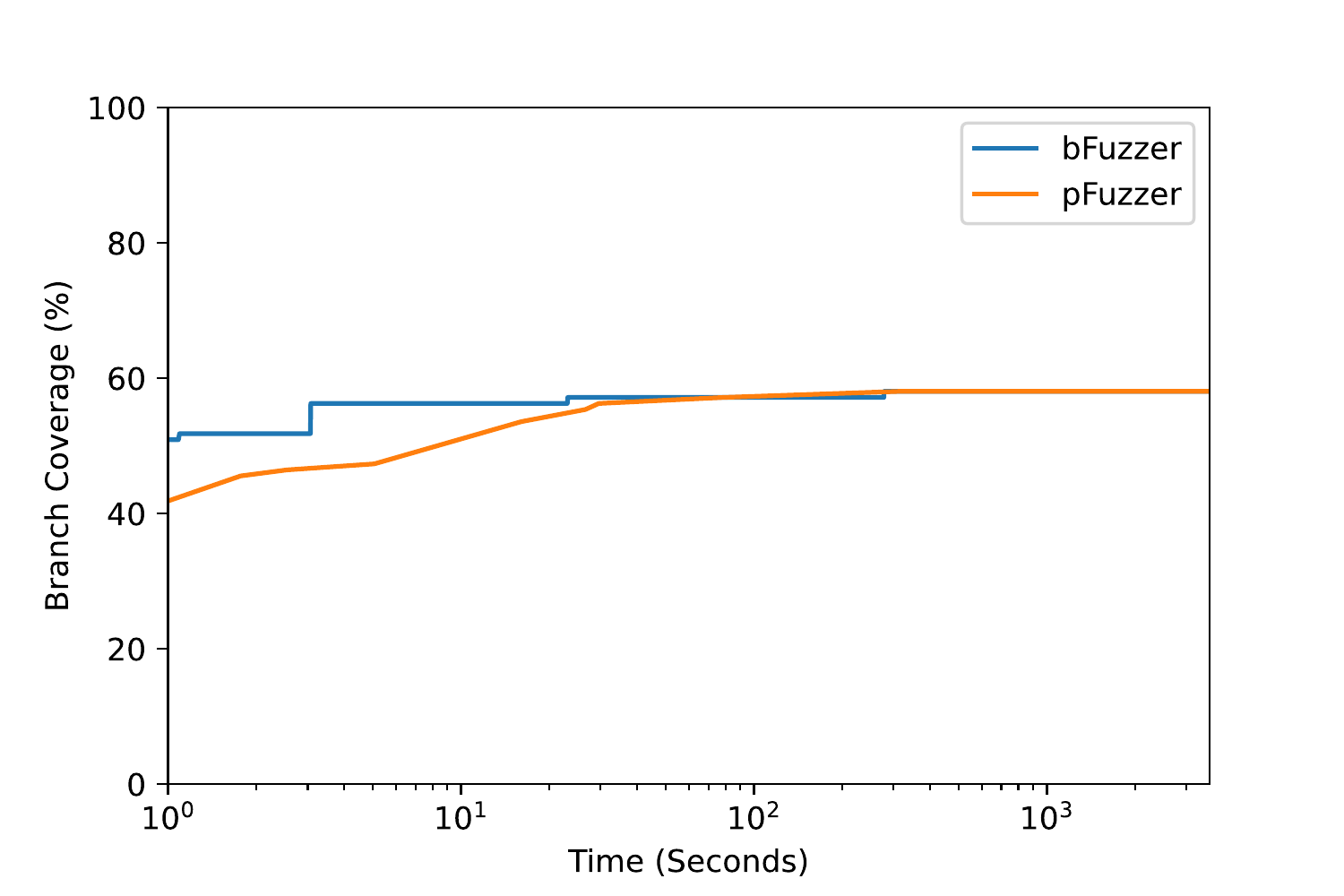}
  \caption{Branch coverage comparison for CSV}
  \label{fig:csvb}
  \end{subfigure}

  \begin{subfigure}[b]{0.35\textwidth}
  \includegraphics[width=\textwidth]{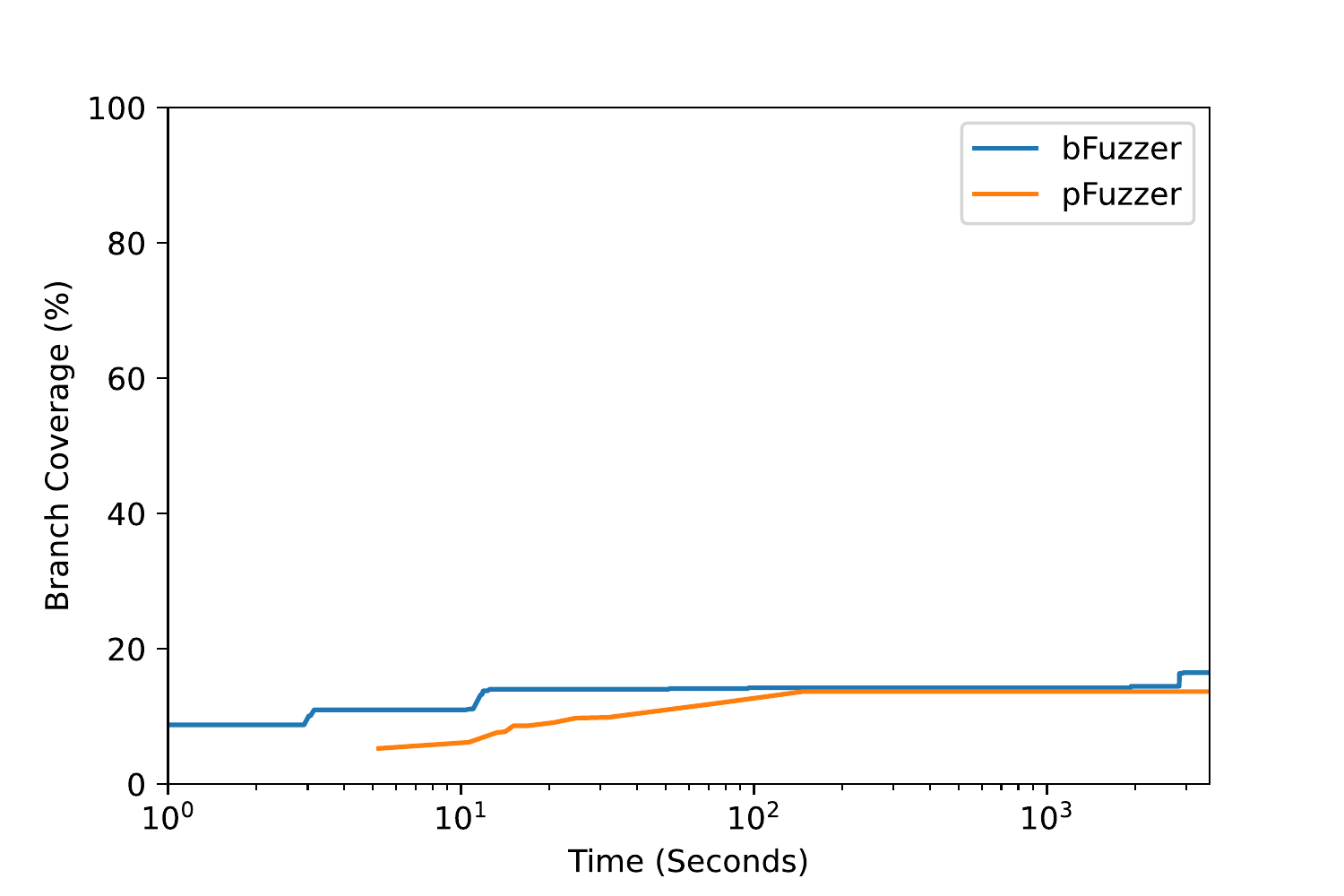}
  \caption{Branch coverage comparison for CJSON}
  \label{fig:cjsonb}
  \end{subfigure}

  \begin{subfigure}[b]{0.35\textwidth}
  \includegraphics[width=\textwidth]{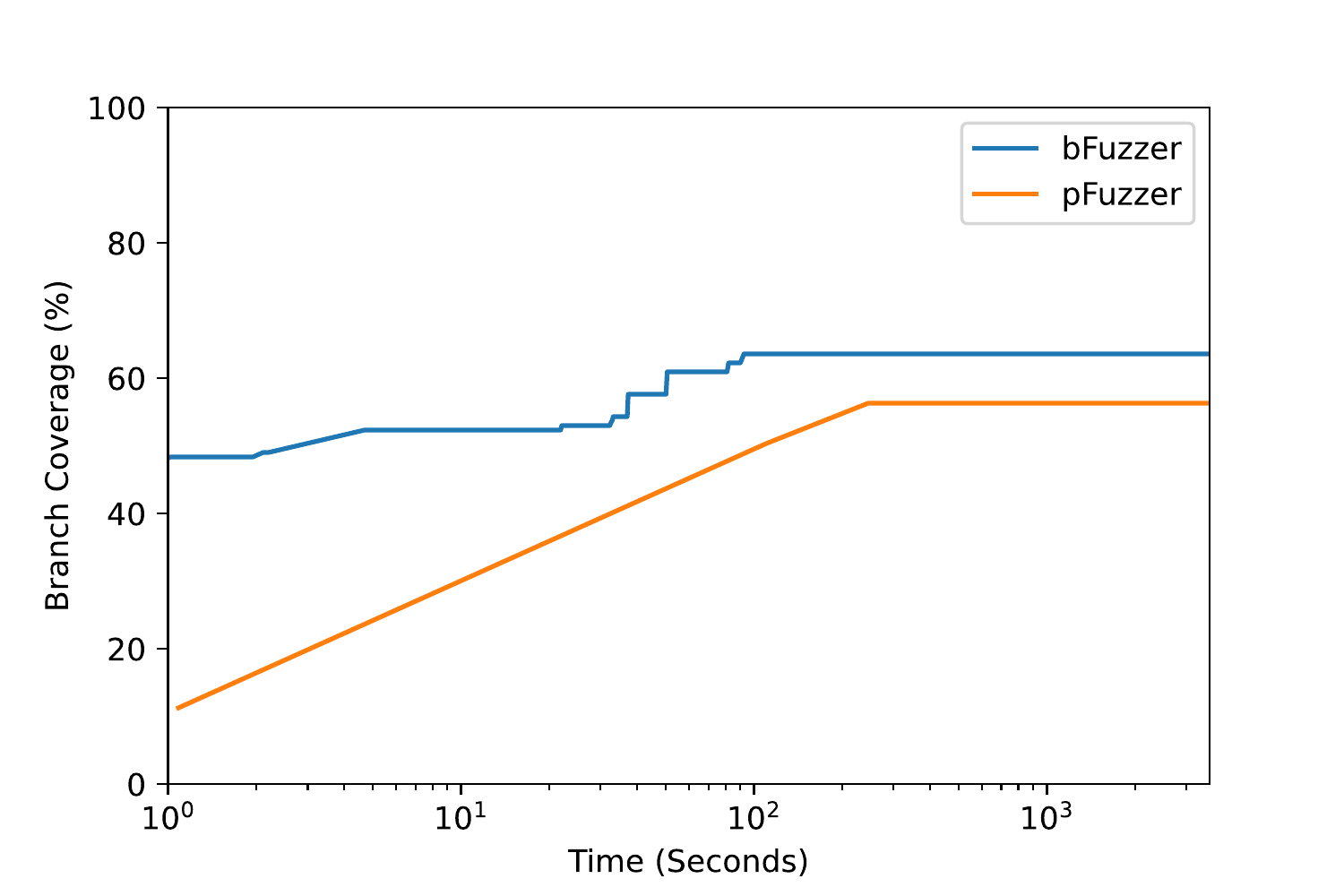}
  \caption{Branch coverage comparison for TinyC}
  \label{fig:tinycb}
  \end{subfigure}

  \begin{subfigure}[b]{0.35\textwidth}
  \includegraphics[width=\textwidth]{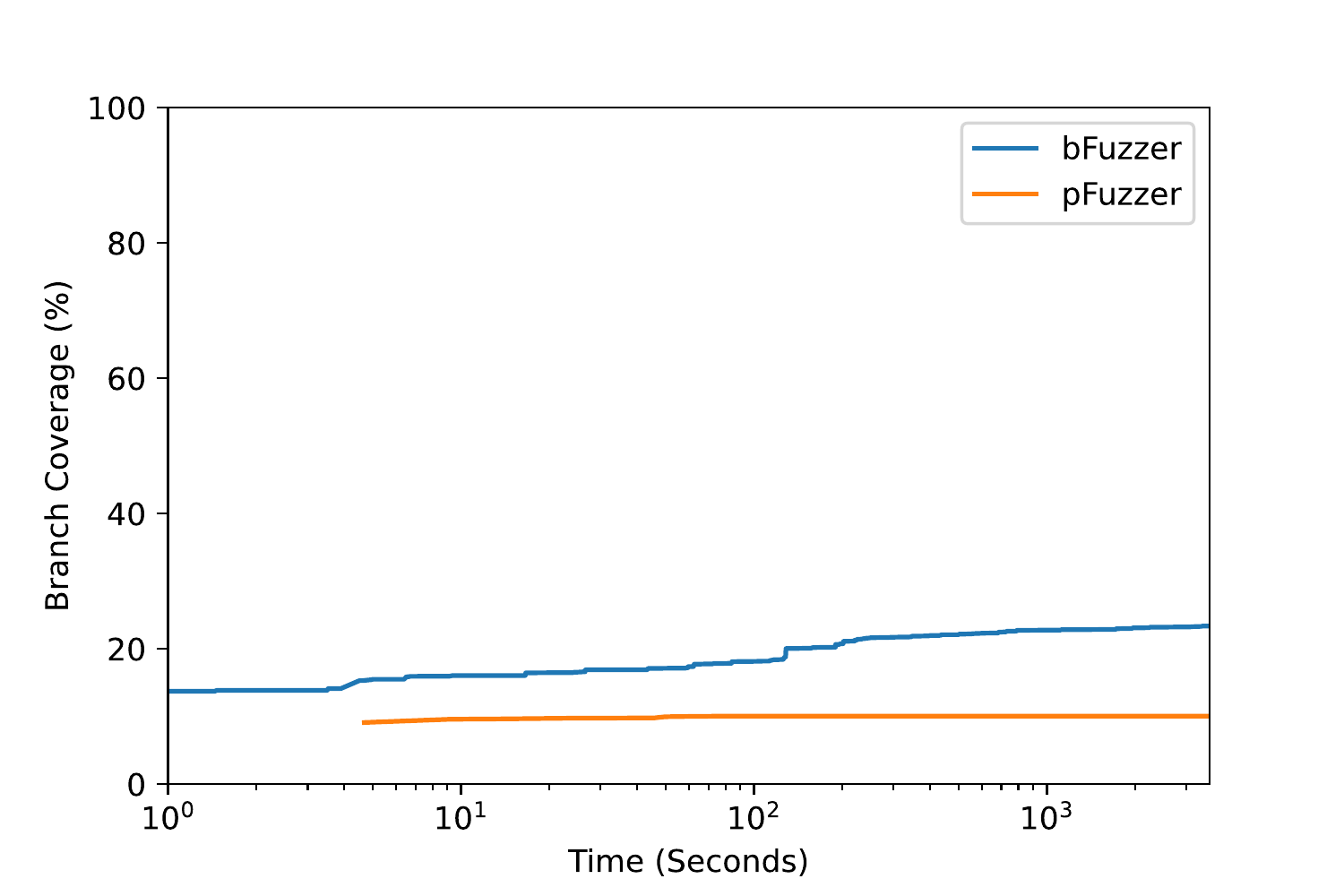}
  \caption{Branch coverage comparison for MJS}
  \label{fig:mjsb}
  \end{subfigure}
  \caption{Evolution of branch coverage in time}
  \label{fig:b}
\end{figure}

Our subjects include programs with simple regular grammars (RG) such as
\ini~\cite{inih} and \csv~\cite{csv} as well as more complex context-free
grammars (CFG) such as \json~\cite{cjson}, \tinyc~\cite{tinyC} (a subset of the C language), and \mjs~\cite{mJS} (a subset of JavaScript).
Given that all of these languages are textual, we used the \emph{printable}
subset ($|\alpha| = 100$ bytes) for evaluation rather than all 256~bytes.
We further set a maximum limit of 1000~bytes per input. That is, any prefix
reaching more than 1000~bytes was discarded, and the computation started
again. This way, we do not get stuck trying to complete
deep unbalanced parentheses.
The programs \json, \tinyc, and \mjs use lexers with slight differences. For \json,
a lexer is used to identify \<true>, \<false> and \<null>. Crucially, the
lexer is used only on locations where these tokens are allowed.
Hence, once the tokens are constructed, there is no backtracking required. For
\tinyc on the other hand, the lexer supplies the tokens to the parser, and
the parser decides whether the provided token is legal or not. Hence,
backtracking is required. For both, we modified the lexer to return correct
feedback using a \emph{trie} that was primed with the known tokens.
\json does not provide extra information on \Rincorrect, while
the \tinyc always provides information on where the parse failure occurred.
Since the \mjs lexer was much more complex than both \tinyc and \json, we left
\mjs without lexer modification.

\pFuzzer on the other hand, does not
require a trie as it is able to recover tokens compared using \<strcmp>
and related library calls. Tries do not work well
with \pFuzzer as they rely on lookups rather than comparisons to determine
whether the input byte is correct.\footnote{Unfortunately the tainting engine of
\pFuzzer crashes when including our trie implementation
to the subject. Hence, we could not run \pFuzzer on the subjects modified
with a trie.} Hence, we do not use tries for producing inputs with \pFuzzer.

Each subject demonstrates a
different facet of the \bFuzzer algorithm. \json demonstrates how using a trie,
we can recover tokens efficiently when there are no further constraints imposed.
The case of \tinyc demonstrates how precise feedback can help when there
are other syntactic constraints imposed on the token. The case of \mjs
demonstrates how backtracking can help. Finally for \ini, any string is
a valid prefix, and for \csv, any string is a valid input.
These demonstrate how the random choice of
a byte is sufficient to cover the input language effectively.

We used the implementation of \pFuzzer that is available from
Mathis et al.~\cite{mathis2019pdfimpl}.
They~\cite{mathis2019parser} use code coverage and input coverage
to evaluate \pFuzzer. In particular, the input coverage is based on the number
of tokens identified. Both \pFuzzer and \bFuzzer generate inputs with similar
tokens, hence it is not a differentiating factor between the
two, and it depends on chance which tokens each tool finds first.

However, there is something even more important. The essential idea of
both \pFuzzer and \bFuzzer is to \emph{fuzz} the system under test. That is,
both parsers need to generate a sufficient number of inputs to exercise the
underlying system. The input parser is merely a gatekeeper, and input
languages such as \json are merely containers for more interesting data.
Hence, these parsers should be judged on the \emph{number} and variety
of \emph{valid inputs} that they are able to produce while exercising the
same inputs features. We note that coverage obtained from valid
inputs is a good proxy for the parser input features exercised. Hence,
we use the number of \emph{unique inputs} produced, and the maximum, and
mean length of inputs produced (along with structural coverage measures)
as metrics to evaluate both fuzzers.

\begin{table*}[ht]
\begin{minipage}{.49\linewidth}
  \captionof{table}{\pFuzzer Evaluation}
\centering
\resizebox{\columnwidth}{!}{%
\begin{tabular}{|l|r|r|r|r|r|}
\hline
  & Unique Valid & Max Len & Mean Len & Line\% & Branch\%\\
\hline
\ini   & 6   & 5   & 4.6   &  75.56 & 56.1 \\
\csv   & 9   & 14  & 6.88  &  67.26 & 58.04 \\
\json  & 13  & 12  & 7.6   &  14.92 & 13.66 \\
\tinyc & 3   & 13  & 11.66 &  72.77 & 56.29 \\
\mjs   & 8   &  13 & 5.5   &  20.96 & 10.02 \\
\hline
\end{tabular}
  }
\label{tbl:pfuzz}
\end{minipage}\hfill
\begin{minipage}{.49\linewidth}
\captionof{table}{\bFuzzer Evaluation}
\centering
\resizebox{\columnwidth}{!}{%
\begin{tabular}{|l|r|r|r|r|r|}
  \hline
  & Unique Valid & Max Len & Mean Len & Line\% & Branch\%\\
\hline
\ini    &   3869   & 216  & 37.23  &  76.3  & 59.76 \\
\csv    &   132514 & 18   & 6.67   &  67.26 & 58.04 \\
\json   &   2222   & 431  & 14.34  &  18.70 & 16.46 \\
\tinyc  &   8061   & 42   & 10.64  &  78.71 & 63.58 \\
\mjs    &   32052  & 959  & 8.93   &  32.42 & 23.35 \\
\hline
\end{tabular}
  }
\label{tbl:bfuzz}
\end{minipage}
\end{table*}

\pFuzzer was run as described by the authors, and the results are provided in \Cref{tbl:pfuzz}.
Next, \bFuzzer was run on the same targets, and the results are provided in \Cref{tbl:bfuzz}.
Both fuzzers were run for one hour. We monitored their coverage behavior and
found that after one hour both fuzzers find new inputs only sparsely. We note that the
original evaluation of \pFuzzer was run for 48 hours. However, we limited our
run to just one hour for two reasons: (1) Both fuzzers, if given enough time,
can achieve saturation in input coverage, (2) Our aim is to show that
the \bFuzzer approach is a reasonable approach when one cannot instrument
the program for any reason. One hour is sufficient to demonstrate this.

We evaluated both tools on a Docker container with a base system configuration of
MacBookPro13,3 with a Quad-Core Intel Core i7
processor with 4 cores running at 2.7GHz.
The L2-Cache was 256 KB, and L3-Cache
was 8 MB. The RAM of the base system was 16 GB. The Docker container
was allocated 7 cores, had 12GB RAM, and ran Ubuntu 18.04.


\subsection{Discussion}
\Cref{fig:b} shows how the branch coverage evolves over time. 
 Table~\ref{tbl:pfuzz} contains the final statistics for \pFuzzer, while
Table~\ref{tbl:bfuzz} contains the final statistics for \bFuzzer.

We see that \bFuzzer is more \emph{efficient} in producing valid inputs than
\pFuzzer. The number of unique inputs produced by \bFuzzer (Table~\ref{tbl:bfuzz})
is multiple orders of magnitude larger than the inputs produced by \pFuzzer
(Table~\ref{tbl:pfuzz}).
\begin{result}
  \bFuzzer can generate unique valid inputs \textbf{many orders faster} than \pFuzzer
  for all programs checked.
\end{result}
Similarly, \bFuzzer is able to produce much larger inputs than \pFuzzer
(maximum length). The mean length of inputs produced by \bFuzzer is also either
larger, or comparable to that produced by \pFuzzer, which indicates a
greater variety.
Finally, \bFuzzer is more \emph{effective} than \pFuzzer for \ini,
\json, \tinyc, and \mjs judging by the coverage obtained.
\begin{result}
  \bFuzzer induced more coverage than \pFuzzer for \ini, \json, \tinyc,  and
  \mjs and equal coverage for \csv during a one hour run.
\end{result}

%


\bFuzzer is a \blackbox fuzzer and hence has limited intelligence
compared to \pFuzzer. So why is \bFuzzer more \emph{efficient} than \pFuzzer in producing
valid inputs? And why is it more \emph{effective} than
\pFuzzer in covering input features (except for \csv)?
One reason may hark back to the primary lesson in fuzzing: When
it comes to the trade off between intelligence and efficiency in generating
inputs, in the absence of overwhelming advantages, efficiency usually
wins~\cite{bohme2014on}.
Secondly, unless there is a strong rational basis for heuristics used, random
methods are best at avoiding bias~\cite{hamlet2006when}.
While the basic idea of \pFuzzer (identify a failure to parse as soon as it
occurs) is certainly worthwhile, we show that a simple trade off for efficiency
in execution --- that of randomly choosing from every option there is, in return
for avoiding instrumentation overhead --- has a disproportionate impact in the
effectiveness.\footnote{
  This is the same reason \pFuzzer performs so well against KLEE.
}

However, there is an important thing to note: \emph{We do not claim that the
\blackbox approach of \bFuzzer is superior to the \whitebox approach of
\pFuzzer}.
The results that we see here are merely due to the constraints of
the particular implementation, which is a research prototype. It is indeed
possible to optimize the \pFuzzer implementation such that the overhead
of instrumentation is limited, and it is likely that such an implementation
would be far faster than \bFuzzer. Indeed, \pFuzzer and its extension,
\lFuzzer have a lot more information to work with, which means that they
can make intelligent trade offs.
Hence, the evaluation is only meant to
showcase the feasibility of the \bFuzzer approach.

However, instrumentation hardly works in these cases: \done{This is great, but how about formulating this in the negative, as in ``Instrumentation hardly works in these cases: The source code of the program is \emph{not} available, the program \emph{cannot} be modified, etc.''? Then this would implicitly also list all the cases in which \bFuzzer would be superior -- AZ}
\begin{enumerate}
\item The source code of the program may not be available.
\item One may not be able to modify the compiled program to insert the instrumentation.
\item The program may be using external libraries or
remote interfaces for validation which are opaque to the instrumentation.
\item The particular instrumentation used by the \whitebox approach
 is unavailable in the particular
language, or the particular compiler cannot be used with the program.
Research prototypes are typically implemented for a single programming
language, usually for a single compiler or framework such as LLVM and are using very specific
  library versions with their own idiosyncracies (for example, pFuzzer instrumentation
    requires specific LLVM versions).
\item The program is multi-threaded, but the
data structures used in the instrumentation are not thread-safe.
\item  Instrumentation adds significant overhead to the program execution.
\item The particular instrumentation used affects the processing.
  For example, the additional overhead of instrumentation may make it
    impossible to observe ephemeral states  (e.g. race conditions), which may
    influence the parsing.
\end{enumerate}
Many of these conditions are often present in the real world.
The main attraction of \bFuzzer is that
it is the \emph{only} alternative so far when one has no access to
instrumentation based feedback.

There are numerous programs that
process files out there\footnote{Wikipedia lists 1,435 input formats~\cite{mathis2019parser}.},
and a significant chunk of these are handcrafted due to concerns of efficiency
and ease of error reporting and recovery.
However, many of these parsers are written in programming
languages with limited support for instrumentation (and indeed the
particular instrumentation that a \whitebox fuzzer may require).
Hence, \blackbox techniques such as ours are especially important in the real world.

\section{Threats to validity}
\label{sec:threats}
Our evaluation is subject to threats to validity.
\begin{description}
  \item [External Validity.] External validity is concerned with
    generalizability of results. This is largely determined by how
    representative our data set is. In our case, the study was conducted
    on five programs, which are of a relatively small size. Hence, we acknowledge
    this threat to validity. A mitigation is that while the programs
    themselves are small, they are well used in the real world. JSON is the
    underlying data exchange format for most of the web, and its parsers
    are known to have inconsistencies~\cite{seriot2016parsing}. Similarly,
    C underlies most modern performance sensitive code, as well as embedded
    systems, while JavaScript is the foundation of the dynamic web. Hence,
    while the programs themselves may not be representative, their input
    languages are representative of the real world.
  \item [Internal Validity.] Internal validity is concerned with the correctness
    of our implementation and our evaluation. Given that our program like
    every other program, is subject to bugs, we acknowledge this threat
    to our empirical evaluation. We have tried to mitigate it by keeping our
    implementation as simple as possible, and ensuring that our program works
    well given small well understood languages with specific properties.
  \item [Construct Validity.] Construct validity is concerned with whether the
    metric we use is actually the right metric for the property we want to
    measure---in our case, whether valid inputs produced
    and the coverage obtained are a reasonable proxy for the effectiveness of
    the fuzzer. Again, we acknowledge this threat to validity, and our
    mitigation is that both coverage and valid inputs produced are common
    metrics used for fuzzer effectiveness measurement.
\end{description}

\section{Related Work}
\label{sec:related}

The three main approaches to fuzzing are \emph{\blackbox} fuzzing,
\emph{specification based fuzzing} and \emph{\whitebox} fuzzing~\cite{godefroid2020fuzzing}.

\subsection{\Blackbox Fuzzing}
\Blackbox fuzzers  operate with little knowledge of the
internals or the specification of the program~\cite{boehme2020fuzzing}. Given
that pure random fuzzing is extremely ineffective when it comes to
complex input languages, modern \blackbox fuzzers
almost always start with a seed corpora of well formed inputs, and mutate
these inputs to generate inputs that are close enough to be valid, but
dissimilar enough to explore new code paths.
The first fuzzer produced by Miller et al.~\cite{miller90an} was a pure random
\blackbox fuzzer. The advantage of \blackbox fuzzers is that they assume very
little about the program in question, and they impose little on the
program runtime (i.e. no feedback).

\subsection{Specification based Fuzzing}
If an input specification is available, one can instead use the specification
based fuzzers such as grammar fuzzers. Such fuzzers rely on the grammar to
generate well formed inputs~\cite{gopinath2019building}. The advantage of such
fuzzers is that they are very efficient when it comes to exploring the
features of the input language, and can generate valid inputs very fast. The
disadvantage is that such a specification has to exist in the first place, and
such specifications, when available, are often obsolete, incomplete, or incorrect.
There are a few tools such as \emph{GLADE}~\cite{bastani2017synthesizing} that can
synthesise the input grammar from sample inputs. However, we need a corpus of
sample inputs that exercise all features of the program in the first place to
use them.

\subsection{\Whitebox Fuzzing}
If access to program source code is available, one can rely on \whitebox methods
such as using symbolic execution frameworks like \emph{KLEE}~\cite{cadar2008klee} and \emph{SAGE}~\cite{godefroid2012sage} to generate
inputs that explore the input space. If in addition to access to the program
source code, the program can also be run under instrumentation, one can make use
of coverage driven fuzzers.


The \whitebox fuzzer \pFuzzer{}~\cite{mathis2019parser} is closely related to \bFuzzer,
and we discuss it in detail next.
Similar to \pFuzzer is \lFuzzer~\cite{mathis2020learning} which adds
heuristics for extracting tokens from parsers.
\emph{BuzzFuzz}~\cite{ganesh2009taint},
\emph{Taint Fuzz}~\cite{liang2013effective} and \emph{Angora}~\cite{chen2018angora} also use
tainting to identify the part of the input that caused the current failure,
and exclusively mutate those bytes.

\subsection{\pFuzzer}
\done{Have a subsection for the earlier approaches, too -- AZ}

A promising approach towards generating inputs for programs with structured
input is the \emph{parser directed fuzzing} strategy (\pFuzzerNS)~\cite{mathis2019parser}.
In this approach, one starts with a random character which is fed to the
program running under instrumentation.
If the program rejects the input (i.e. returns a non-zero exit code), the comparisons
made on the randomly chosen input character are used to substitute the compared character
with one of the values it was compared to. \pFuzzerNS
uses branch coverage guidance (i.e. substitutions on inputs that covered
new code are preferred) to explore all possible substitutions.
Whenever the program returns with exit code zero and new branches
are covered compared to the already reported valid inputs,
\pFuzzerNS~reports this input as valid and marks the covered branches as covered.
For example, if only digits were compared against the last character, a random
digit is used to replace the last character, and the process repeats with the input
that covered the most new branches until then. Proceeding in this fashion,
\pFuzzer generates longer and longer sequences of valid prefixes which
ultimately produce valid inputs.
Mathis et al. extends this approach in \lFuzzer~\cite{mathis2020learning} which
includes additional heuristics for identifying tokens which are valid at
any given point. Since this includes
more assumptions about the code, it is no longer comparable to \bFuzzer.
Hence, we do not use \lFuzzer as the baseline.

\pFuzzer has a number of limitations that are mitigated by \bFuzzer.

\subsubsection{Table driven parsers}
Comparisons cannot be typically mined from table driven parsers, as the
bytes read are not compared against but rather used to look up
a table entry. We note that \bFuzzer does not have a problem against table
driven parsers (and indeed, as a \blackbox fuzzer, one may not even know what
kind of a program we fuzz, and hence obtaining the grammar for it may be moot).

\subsubsection{Tokenization}
Lexers cause a break in the data flow. That is, given this fragment,
\begin{lstlisting}[style=python]
if next_char() == '(': return LPAREN
\end{lstlisting}
LPAREN no longer has a (direct) taint (the data flow is broken), and hence is not tracked. However, \bFuzzer
does not have a problem with this kind of code so long as there is precise failure feedback.

\subsubsection{Can one trust the mined comparisons?}
Comparisons may be used for either the \emph{continue parsing} branch or the \emph{exit with failure} branch.
Consider the fragment below. Here, the byte at index \<c> is compared to uppercase values, and \emph{if it is} an uppercase value, \Rincorrect is returned.
\begin{lstlisting}[style=python]
if input[c] in str.upper: raise Incorrect()
parse_further(input, c+1)
\end{lstlisting}
Now, consider the fragment below.
\begin{lstlisting}[style=python]
if not (input[c] in str.lower): raise Incorrect()
parse_further(input, c+1)
\end{lstlisting}
Here, we check if the byte at index \<c> is a lowercase value, and \emph{if it is not}, it
returns \Rincorrect. Hence, there is no single pool of bytes (either compared bytes or its complement)
that can be consistently used to replace the byte compared, and hence advance the parsing.

Since \pFuzzer relies on compared characters as the pool from which to choose the next character,
this represents a blind spot for \pFuzzer. \done{Sorry to open this again but when looking at it now, I thinks it still does not apply as the character c would be a random character at first. Hence the comparison with [ would be seen. And even if it would by a slim chance take 0 as the first option, \pFuzzer would at some point come again to this position (with another prefix for example), and the likelihood that it again randomly chooses 0 is even slimmer. -- BM; see if the new explanation makes sense -- RG}
Since \bFuzzer does not use comparisons, this is not a problem for \bFuzzer.

\subsubsection{Semantic restrictions}
In many cases, the comparison of an input to a valid value comes after the data
has been converted from string to another data type. In the below expression,
the check is made to the integer value, parsed from say ` 0b11', which results
in `3', and byte taints are no longer applicable to the new integer value.
\begin{lstlisting}[style=python]
ival = parse_binary_digit(input[2:])
if ival != 3: raise Incorrect()
\end{lstlisting}
As in the tokenization, \bFuzzer does not have a problem with this if either
precise failure feedback is available, or the compared value is only a few bytes
long (so that it can be found by backtracking).

\subsubsection{Recursion}
For the basic algorithm of \pFuzzer, recursion can result in degradation of
performance. For example, considering a simple parenthesis language with
well balanced \term{(} and \term{)}, the probability of closing such
a prefix after 100 steps is about 1\%. That is, ($\frac{1}{n+1}$), and continues
to decrease as more characters are added. \pFuzzer has to rely on a number
of heuristics (which are limited by definition) to avoid this.

While \bFuzzer suffers from the same constraint, as it is faster than
\pFuzzer, \bFuzzer is able to generate a much larger number of valid
prefixes. Hence, the number of valid inputs produced is larger.

\emph{Hybrid fuzzing:}
Both \pFuzzer and \bFuzzer are \emph{valid prefix} based fuzzers,
and can switch back
and forth to each other with little effort while fuzzing the same program.
This can be useful especially considering that \lFuzzer which is an extension
of \pFuzzer has an advantage when it comes to tokens (in that it is able to
identify the valid tokens at any position fast), and \bFuzzer has an advantage
in finding continuations fast when precise feedback is available. Hence,
combining them may produce a better fuzzer.

\subsection{Exhaustive Testing}

Bounded Exhaustive Testing~\cite{howden2008adequacy,jagannath2009reducing,coppit2005software} is a
testing technique where the properties of a program under test are verified by
exhaustively testing up to a certain depth. Exhaustive testing is a powerful
technique, and Goodenough et al.~\cite{goodenough1975toward} points out that
the exhaustive technique can serve as a proof. It can also introduce oracles with
higher order logic such as $\forall$, $\exists$. That is, it becomes possible
to say that there do not exist two floating point expressions that evaluate
to the same value, or a particular grammar does not have an ambiguous parse
for strings of a fixed depth, or for all functions, a particular approximation
is within bounds.
Given the power of BET, it is under active research~\cite{ahmet2017bounded,rosner2014bounded,dewey2020mimis,aguirre2011incorporating}

Given the utility of BET, any approach that can
reduce the cost of BET is welcome for safety critical application verification.
The fast feedback fuzzing by \bFuzzer can reliably prune a large part of the
input quickly, and do that without running the program under instrumentation
(which can incur an overhead). Hence, \bFuzzer can be immediately used for BET,
and can validate the properties of more programs than previously possible.

\subsection{Cryptanalysis}
The concept of \emph{decoding}~\cite{edgar2017effective} where the lock is
solved one tumbler or position at a time is well known in the
lock picking community,\done{Explain what decoding is. -- AZ} and the importance of side channels are again well
understood in the cryptanalysis community. Our approach is a variation of
\emph{decoding} number locks, and shows the deep synergy between these
fields.
\done{While being a fun side-note, this should be the first thing to be thrown out if we need space :D;
I disagree. This has to stay -- AZ}

\section{Limitations and Future Work}
\label{sec:limitations}
While our approach is better than the state of the art in quickly generating
valid inputs that cover all input features in
\blackbox settings,\done{Given that 2019--2020, we have seen 100 new fuzzer papers, this may be found questionable. Better: Our approach is the only one that quickly generates ... in \emph{black-box} settings ... -- AZ}
\bFuzzer still has a number of
limitations. 

\subsection{Dependence on immediate failure feedback}
The efficiency of \bFuzzer is somewhat dependent on immediate failure feedback.
\bFuzzer works best if the program provides immediate feedback after
each byte whether the byte was expected or not. If this expectation is
not met, \bFuzzer can continue with a \emph{linear degradation} of its
effectiveness, requiring $\frac{256L}{C} \times \sum{C}$ executions in comparison to
just $256 \times L$ executions if the program provides failure feedback only after
reading $C$ bytes (on average).

However, there may be further optimizations possible. For example, if one
can predict the next byte, perhaps through statistical models, especially using
reinforcement learning with hidden
state~\cite{sennhauser2018evaluating, mccallum1996hidden,hasinoff2002reinforcement}
one may be able to reduce the overhead even futher by predicting the next byte.

%

\subsection{Modifying the program for fast failure feedback}
Given that instrumentationless fuzzing may be useful in many scenarios, and
also given that our technique is actually faster than the next best technique
using instrumentation, it may be worthwhile to explore how to apply \bFuzzer to
more programs, by getting them to provide immediate and accurate failure
feedback. In combination with some static analysis, one may incorporate further
information to the return values, or rearrange validation steps so that
validation of earlier bytes come earlier, or splitting the comparisons similar
to the AFL compare-transform pass~\cite{lafintel2020circumventing}
or incorporate the \emph{trie} data structure to common lexer libraries for
accurate failure feedback. These can
likely improve the efficiency of \bFuzzer, and will be a topic for future
research.

\subsection{Dependence on the speed of execution of the program}
As with other fuzzers (similar to, but more so than \pFuzzer), \bFuzzer is
completely reliant on the program under fuzzing being fast to execute. If the
program under consideration is not speedy enough in quickly executing and
returning failures, \bFuzzer will not be efficient. This can be worked
around to some extent by side-channels such as time to process if such
information is available. For example, if we know that processing some byte
takes a fixed amount of time, and taking more time can be a hint that the
program input is incomplete rather than incorrect, one may use this information
in combination with timeouts. Hence, this will be an area of active research for
the future.

\subsection{Information about the program and side channels}
One may reduce the number of executions required by half if one can restrict
the bytes required to only printable ASCII letters. Similarly, one may avoid
combinatorial explosion during backtracking if one knows beforehand the tokens
used by the program, or even a large dictionary of words likely to be used as
tokens, or possible skeletal structure of the input required. Further, any side
channel about how the program processes its input may be incorporated, and could
make fuzzing faster and more efficient. How to do this without requiring
active instrumentation will be a future focus.

\subsection{Pairing with a grammar miner}
We mentioned using learners to predict the next byte previously. One may also
approach this in a more direct fashion. One may simply pair \bFuzzer with
a \blackbox grammar miner such as GLADE~\cite{bastani2017synthesizing} and
identify the input specification completely. \bFuzzer would be especially
complementary to GLADE as GLADE requires a few valid samples for it to learn
the input specification, which \bFuzzer can supply.


Further, given a grammar miner that can predict the probability of acceptance of
the next byte, one may compose refute and validate hypotheses on the fly,
improving the efficiency of both the fuzzer as well as the miner, achieving
more than what each are capable of independently.
\done{Another limitation: pFuzzer relies completely on exploration with no exploitation. We can add an exploitation element to make it more effective. Meaning, we can use the already generated valid inputs to our advantage. As a suggestion, in some iterations we can take a previous promising input and truncate it to half then start generation from there. --- indeed, this can
probably go in the future work if we have space. Since we do not do that now
(and have no time to do it) we should not claim it as a difference with
pFuzzer (RG); Also: is this really a problem of the pFuzzer approach? In theory one can make use of a prefix (and lFuzzer already has such a command line flag),
hence this is just an implementation limitation, something we should not claim as a general disadvantage compared to bfuzzer.; I think Bjoern's explanation makes sense.}

\section{Conclusion}
\label{sec:conclusion}
Traditional feedback driven fuzzers rely on running the program under
instrumentation, which reduces the speed of execution of the program. However,
for many real world systems, running a program under instrumentation is
infeasible due to limitations in access, external libraries or 
language used. Hence, an efficient \blackbox approach is needed.

Traditional \blackbox techniques (and even most \whitebox techniques) fare
poorly when it comes to programs with complex input specifications. The
problem is that random generation of inputs, while good at producing
unexpected inputs, are exponentially bad at producing defined values such
as magic bytes and keywords, requiring $256^L$ attempts to produce a
keyword of length $L$. Failure feedback fuzzing is the first approach that
successfully marries the wide applicability of \blackbox fuzzing with
the effectiveness of \whitebox fuzzing when it comes to programs
with complex input specifications.

Our approach relies on the observation that most programs provide accurate
failure feedback to the user on processing a given input, which can be used
to guide the fuzzer. Our \bFuzzer prototype can quickly generate valid inputs
that can cover all input features in an unbiased manner. While side channel
attacks are a common fare in the cryptanalysis community, ours is the first
work marrying side channel information to fuzzing, showing the deep synergy
between the two fields.

We evaluate our fuzzer on programs that require complex inputs such
as \ini, \csv, \json, \tinyc, \mjs, and show that our fuzzer is efficient and
effective in generating valid inputs quickly, and is even more effective than
\pFuzzer which is a \whitebox technique.

Our complete implementation and experiments are available as a Jupyter notebook
\begin{center}
  \footnotesize{\href{https://github.com/vrthra/bFuzzer/blob/master/BFuzzer.ipynb}{https://github.com/vrthra/bFuzzer/blob/master/BFuzzer.ipynb}}
\end{center}


\printbibliography
\end{multicols}
\end{document}